\documentclass[%
 reprint,
 amsmath,amssymb,
 aps,
]{revtex4-1}

\usepackage{graphicx}
\usepackage{dcolumn}
\usepackage{bm}
\usepackage{multirow}
\usepackage{color,soul}

\begin{document}


\title{Mapping rules from Nodal Line Semimetal to Topological Crystalline Insulator in Face centered Cubic Lattice}

\author{Ikuma Tateishi}
 \affiliation{Department of Physics, The University of Tokyo, Bunkyo, Tokyo 133-0033, Japan}
 \email{i.tateishi@hosi.phys.s.u-tokyo.ac.jp}

\date{\today}

\begin{abstract}
We study what kind of topological crystalline insulator phase emerges from nodal line semimetal phases in the face-centered cubic system which is not indicated by topological indices when spin-orbit coupling (SOC) is introduced. We construct an effective model which hosts two different nodal lines phases, and calculated mirror Chern numbers in it by introducing SOC. As a result, we find that the two nodal line phases with different nodal line configurations are mapped to different topological crystalline insulator phases. This result shows that turning off SOC and checking the nodal line configuration can distinguish the two topological crystalline insulator phases, which have not been distinguished by previous methods.

\end{abstract}

\pacs{Valid PACS appear here}
\maketitle



\textit{Introduction.}---
The development of an easy and simple diagnostic method of topological insulators is one of the greatest goals in the field of topological materials science \cite{FuKane,TIclass1,TIclass2,SBI,TQC,CF1}. Nowadays, an advanced diagnostic method called the symmetry-based indicator has been proposed \cite{SBI}. The symmetry-based indicator is given by checking irreducible representations (irreps) on the high-symmetry point (line, plane) of Brilloiun zone (BZ), and it works as a diagnostic method for topological insulators with spin-orbit coupling (SOC) \cite{egSBI1,egSBI2,egSBI3}.

Not only for topological insulators, a diagnostic method for topological semimetals without SOC has also been studied \cite{CF2,YKim,NLZ2Charge}. It is partially because the topological semimetals without SOC are considered to be the "parents states" of many topological states when SOC is introduced \cite{KaneMele1,KaneMele2,Weyl1,Weyl2,SCZ}. The topological semimetals without SOC are divided into two groups. One of them is the semimetals with nodes on the high-symmetry line (plane) in BZ \cite{NLSM2,NLSM3,NLSM4,NLSM5,NLHirayama,ITSnSe,Weyl2}. The other is the semimetals with nodes in the generic points \cite{CF2}. The former is relatively easy to find because the first-principles calculation usually calculates the band dispersion along the high-symmetry lines \cite{DFTsym}. On the other hand, the latter is difficult to find, and thus diagnostic methods for them are energetically studied. As a diagnostic method for them, a symmetry-based method has been proposed, in which the concept of the symmetry-based indicator is used \cite{CF2}. Since the methods use the same concept with the symmetry-based indicator, a "mapping" between them has been discussed, i. e., what kind of topological insulator emerges from a topological semimetal ("parent state") when SOC is introduced \cite{CF2}. However, this mapping is not understood well for the former group, which is topological semimetals with nodes on the high-symmetry lines.

To consider the mapping from topological semimetals with nodes on the high-symmetry lines to topological insulators, there is a suggestive example of a nodal line semimetal, which is a kind of topological semimetals. For inversion and time-reversal(TR) symmetric nodal line semimetals without SOC, a diagnostic method called the $\mathbb{Z}_2$ index is defined \cite{YKim}. This $\mathbb{Z}_2$ index is defined in the same way as the well-known Fu-Kane $\mathbb{Z}_2$ index for TR protected topological insulators \cite{FuKane}. This means that a nodal line semimetal with a non-trivial $\mathbb{Z}_2$ index is mapped to a topological insulator with the same non-trivial $\mathbb{Z}_2$ index, when SOC is introduced \cite{NLSM3,NLSM4}. A recent progress extended the $\mathbb{Z}_2$ index to $\mathbb{Z}_4$ index, and revealed a new classified $\mathbb{Z}_4=2$ phase correspond to nomopole-charged nodal line when SOC is neglected \cite{NLZ4}. It is also revealed that the $\mathbb{Z}_4=2$ monopole nodal line phase is mapped to the higher-order topological insulator phase \cite{HOTI,HOTIBi}. Meanwhile, some materials are proposed as nodal line semimetals with trivial $\mathbb{Z}_4$ index \cite{NLHirayama,ITSnSe}. A question then naturally presents itself: to what kind of topological insulator phase these nodal line semimetals are mapped?

In this paper, to answer the question above, we discuss a mapping from a nodal line semimetal with trivial $\mathbb{Z}_4$ index to topological crystalline insulators \cite{TCI,TCI2}. Based on the proposed material examples in the face-centered cubic lattice (FCC, Space group \#225) system \cite{NLHirayama,ITSnSe}, we construct a generalized two-bands effective model, which keeps the same nodal line structure as the material examples. By using the model, we show that the nodal line in the system is mapped to a mirror Chern number when SOC is introduced. Furthermore, we show that the difference in the configuration of the nodal line corresponds to the difference between the topological crystalline insulator phases. After the calculation and discussion on the model, we show that the phase mapping is actually observed in some materials examples. Since this result distinguishes two topological crystalline insulator phases which can not be distinguished by previous methods, it can lead to a subdividing diagnostic method beyond the symmetry-based indicator.

\textit{Target system.}---
\begin{figure}
    \centering
    \includegraphics[width=8cm]{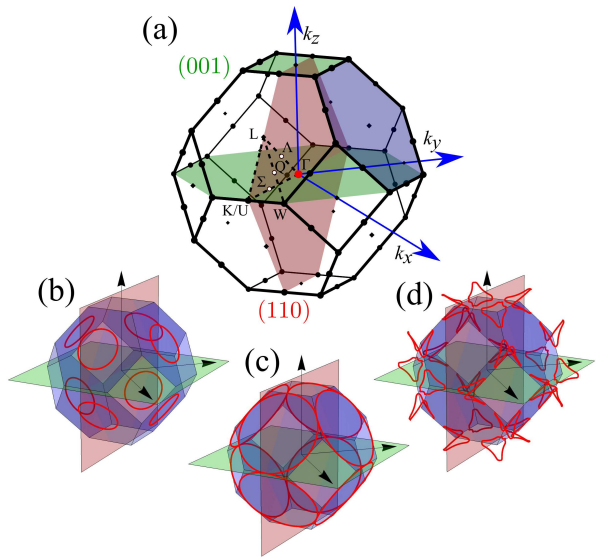}
    \caption{(a) Brillouin zone of face-centered cubic lattice (Space group \#225, $Fm\bar{3}m$). (b)(c)(d) Nodal lines (red lines) in the system we assume. The configuration of nodal lines depends on a parameter. The (001) mirror plane and the (110) mirror plane is shown as green and red planes, respectively. The hexagonal surface of the BZ is shown as a blue plane for convenience, though it is not a high-symmetry plane.}
    \label{fig:nlbz}
\end{figure}
We assume a nodal line semimetal in the FCC system without SOC (Fig.\ref{fig:nlbz}). To focus on a more concrete system, we construct a half-filled two-bands simple effective model based on some realistic materials such as FCC Ca, Sr, Yb, and SnSe, which are proposed as nodal line semimetals \cite{NLHirayama,ITSnSe}. To describe the realistic case well and obtain a generalized model, the two-bands model should satisfy the following assumptions. \\
(i) The nodal lines are given as a result of band inversion at the L-point and they are protected by TR+inversion symmetry \cite{nodal,protnode} (Fig.\ref{fig:nlbz}(b)). \\
(ii) The band inversion at the L-point does not make a mirror protected nodal line on the (110) plane, i. e., the inverted two bands have the same mirror eigenvalues. \\
(iii) By tuning a parameter, nodal lines touch each other and reconnection occurs on the $\Sigma$-line (Fig.\ref{fig:nlbz}(c)). \\
(iv) After the reconnection, the nodal lines are located around the W-point and there is no nodal line on the (001) plane (Fig.\ref{fig:nlbz}(d)). \\
To satisfy the assumption (i), the band structure of the two-bands model must have a crossing point on the $C_2$ invariant line (L-W-line or Q-line), and the two band must have different $C_2$ rotational eigenvalues. Considering these assumptions and the compatibility relation \cite{herring,BC}, the irreps of two bands are decided, $\Gamma_1^+$ and $\Gamma_2^-$ at the L-point, $\Gamma_1$ and $\Gamma_4$ at the W-point, and $\Gamma_1$ and $\Gamma_1$ on the $\Sigma$-line (See Appendix \ref{SMsec:modelirrep} for the detail). Since the irreps of bases have been obtained, we construct two by two models with nodal line by using the $\bm{k} \cdot \bm{p}$-perturbation within the second-order of $k$ for each high-symmetry point (line). Here we call the constructed model as local models at the L-point, the W-point, and the $\Sigma$-line. When SOC term with an amplitude $\chi~(>0)$ is introduced into the local models, they get gaped and the Berry curvature and the mirror Chern numbers can be calculated. By considering a small SOC (small $\chi$) case, we discuss a "mapping rule" from nodal lines to the mirror Chern numbers. Especially on $\Sigma$-line, we discuss how the reconnection of nodal lines affects the Berry curvature and the mirror Chern numbers.

\begin{figure*}
    \centering
    \includegraphics[width=17cm]{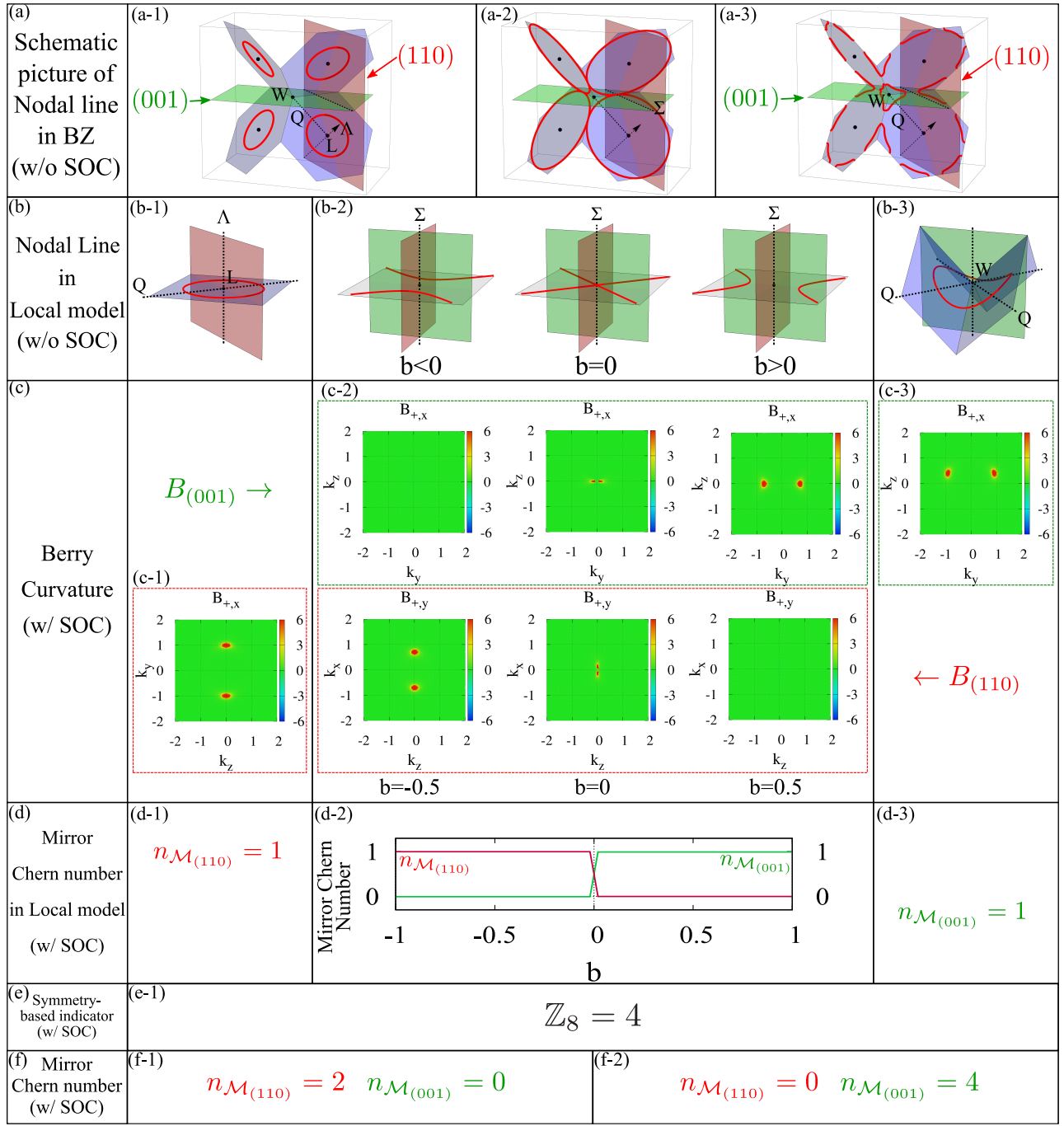}
    \caption{(a) Schematic pictures of nodal lines (red ring) in the target systems. (a-1) Nodal lines around the L-point. The nodal lines penetrate the (110) mirror plane (red plane). (a-2) Reconnection of nodal lines. By tuning a parameter, the radius of the nodal line is changed and the reconnection of nodal lines occurs on the $\Sigma$-line. (a-3) Nodal lines around the W-point. (b) Schematic pictures of the local model and nodal lines. (b-1) Local model around the L-point. (b-2) Local model around the $\Sigma$-line. The left, middle, and right ones are the $b<0$, $b=0$, and $b>0$ case, respectively. (b-3) Local model around the W-point. (c) Momentum dependence of the Berry curvature. (c-1) (110) component of Berry curvature in the local model around the L-point. (c-2) Upper three panels are the (001) component and lower three panels are the (110) component of the Berry curvature in the local model around the $\Sigma$-line. For both rows, the left one is the $b<0$ case, the middle one is the $b=0$ case, and the right one is the $b>0$ case. (c-3) (001) component of Berry curvature in the local model around the W-point. (d) Mirror Chern number of the local models. (d-1) (110) mirror Chern number in the local model around the L-point. (d-2) $b$ dependence of the (110) mirror Chern number and the (001) mirror Chern number in the local model around the $\Sigma$-line. (d-3) (001) mirror Chern number in the local model around the W-point. (e) Symmetry-based indicator. The indicator is the same for both phases. (f) Mirror Chern numbers of the whole BZ. These values are calculated from (d) by counting how many L-points (W-points) exist on the (110) mirror plane ((001) mirror plane).}
    \label{fig:main}
\end{figure*}

\textit{Around the L-point.}---
First, we discuss the local model at the L-point (The detail of the calculation is given in the Appendix \ref{SMsec:modelder}, \ref{SMsec:MCN}). The local model without SOC is written as
\begin{equation}
    H_L = k_z \sigma_x + (k_x^2 + k_y^2 + k_z^2 - \Delta^2) \sigma_z .
\end{equation}
Here the $k_z$ direction is parallel to the $\Lambda$-line, the (110) mirror invariant plane is the $k_x=0$ plane, and $\Delta$ is a real constant. The $\sigma_{0,x,y,z}$ are the Pauli matrices for orbitals. In this model, a ring of nodal line emerges on the $k_z=0$ plane (Fig.\ref{fig:main}(b-1)). By introducing a Rashba type SOC term with an amplitude $\chi~(>0)$ \cite{Rashba1,Rashba2}, the four by four local model with SOC is written as
\begin{equation}
    H_{L,\mathrm{soc}}= H_L s_0 + \chi \sigma_y (-k_y s_x + k_x s_y) ,
\end{equation}
where the $s_{0,x,y,z}$ are the Pauli matrices for the spin degree of freedom. As a result of introducing the SOC term, the nodal line vanishes and the model becomes gaped. On the $k_x=0$ plane, this model can be block diagonalized for the mirror eigenvalue. The block with $+i$ mirror eigenvalues is
\begin{equation}
    H_{L,+} = k_z \sigma_x - \chi k_y \sigma_y + (k_x^2 + k_y^2 + k_z^2 - \Delta^2) \sigma_z .
\end{equation}
The $x$ component of the  Berry curvature of the occupied band is given as
\begin{equation}
    B_{L,+,x} = \frac{\chi}{2 R_L^3} (k_y^2 + k_z^2 + \Delta^2),
\end{equation}
where $R_L$ is defined as
\begin{equation}
    R_L = \sqrt{ k_z^2 + \chi^2 k_y^2 + (k_y^2 + k_z^2 -\Delta^2)^2} .
\end{equation}
The mirror Chern number of this local model is calculated as $n_{{\cal M}_{(110)}}=1$ (Fig.\ref{fig:main}(d-1)). In small $\chi$ limit, the Berry curvature has sharp peaks on the points where nodal lines penetrate when SOC is neglected (Fig.\ref{fig:main}(c-1)). Since the mirror Chern number is topological invariant, it must be kept in a large $\chi$ case. Therefore, the nodal line is considered as a source of the mirror Chern number. There are two non-equivalent L-point on the (110) plane, and thus the mirror Chern number of the whole BZ is  $n_{{\cal M}_{(110)}}=2 $ (Fig.\ref{fig:main}(f-1)). Here "non-equivalent" means that the two L-points are not connected by the reciprocal lattice vectors.

\textit{Around the W-point.}---
Next, we discuss the local model at the W-point (The detail of the calculation is given in the Appendix \ref{SMsec:modelder}, \ref{SMsec:MCN}). The local model without SOC is written as 
\begin{equation}
    H_W = (k_z  + a k_x^2 - a k_y^2) \sigma_x + (k_x^2 + k_y^2 + k_z^2 - \Delta^2) \sigma_z .
\end{equation}
Here the $k_z$ direction is parallel to the $C_2$ invariant line on the (001) mirror invariant plane ($k_x=0$ plane) and $\Delta$ and $a$ are real constants. In this model, a nodal line emerges around the W-point and it is oscillating in the $k_z$ direction, keeping $C_2$ rotation symmetry around the $k_z$ axis (Fig.\ref{fig:main}(b-3)). The local model with SOC is
\begin{equation}
    H_{W,\mathrm{soc}}= H_W s_0 + \chi \sigma_y (-k_y s_x + k_x s_y) .
\end{equation}
On the $k_x=0$ plane, this model can be diagonalized and the block of $+i$ mirror eigenvalues is
\begin{equation}
    H_{W,+} = (k_z - a k_y^2) \sigma_x - \chi k_y \sigma_y + (k_y^2 + k_z^2 - \Delta^2) \sigma_z .
\end{equation}
The $y$ component of the Berry curvature of the occupied band is 
\begin{equation}
    B_{W,+,x} = \frac{\chi}{2 R_W^3} (k_y^2 + k_z^2 + \Delta^2 + 2 a k_y^2 k_z) ,
\end{equation}
where $R_W$ is defined as
\begin{equation}
    R_W = \sqrt{ (k_z - a k_y^2)^2 + \chi^2 k_y^2 + (k_y^2 + k_z^2 -\Delta^2)^2} .
\end{equation}
The mirror Chern number of this local is calculated as $n_{{\cal M}_{(001)}}=1$ (Fig.\ref{fig:main}(d-3)). The sharp peak feature of the Berry curvature is also seen in this case (Fig.\ref{fig:main}(c-3)), and thus the nodal line is considered as a source of the mirror Chern number. There are four non-equivalent X-point on the (001) plane and the mirror Chern number of the whole BZ is calculated as $n_{{\cal M}_{(001)}}=4$ (Fig.\ref{fig:main}(f-2))

\textit{Around the $\Sigma$-line}---
Finally, we discuss the local model on the $\Sigma$-line (The detail of the calculation is given in the Appendix \ref{SMsec:modelder}, \ref{SMsec:MCN}). The local model without SOC is written as
\begin{equation}
    H_\Sigma = k_z \sigma_x + (k_x^2 - k_y^2 + k_z^2 + b) \sigma_z .
\end{equation}
Here the $k_z$ direction is parallel to the $\Sigma$-line and $b$ is a real tunable parameter. The (001) mirror plane is the $k_x=0$ plane, and the (110) mirror plane is the $k_y=0$ plane. In this model, hyperbolic nodal lines emerge on the $k_z=0$ plane. The nodal lines penetrate the $k_x=0$ plane for a $b<0$ case, and the $k_y=0$ for a $b>0$ case (Fig.\ref{fig:main}(b-2)). When $b=0$, the two nodal lines touch each other, and this touching corresponds to the reconnection of nodal line (Fig.\ref{fig:main}(a-2)). The local model with SOC is
\begin{equation}
    H_{\Sigma,\mathrm{soc}}= H_\Sigma s_0 + \chi \sigma_y (-k_y s_x + k_x s_y) .
    \label{eq:SlineSOC}
\end{equation}
In this model, there are two mirror invariant planes, and thus we can discuss how the reconnection of nodal lines affects the mirror Chern numbers.
First, we focus on the $k_x=0$ plane. The $+i$ mirror eigenvalue block is
\begin{equation}
    H_{\Sigma,+}^{(k_x=0)} = k_z \sigma_x - \chi k_y \sigma_y + (-k_y^2 + k_z^2 + b) \sigma_z .
\end{equation}
The $x$ component of the Berry curvature of the occupied band is
\begin{equation}
    B_{\Sigma,+,x}^{(k_x=0)} = \frac{-\chi}{2 R_{\Sigma,(k_x=0)}^3} (-k_y^2 + k_z^2 - b) ,
\end{equation}
where the $R_{\Sigma,(k_x=0)}$ is defined as
\begin{equation}
    R_{\Sigma,(k_x=0)} = \sqrt{ k_z^2 + \chi^2 k_y^2 + (-k_y^2 + k_z^2 + b)^2} .
\end{equation}
When $b>0$, the Berry curvature $B_{\Sigma,+,x}^{(k_x=0)}$ has sharp peaks (Fig.\ref{fig:main}(c-2),upper right) at the points where the nodal lines penetrate when SOC is neglected. As $b$ getting smaller, the two peaks get closer, and they meet each other when $b=0$ (Fig.\ref{fig:main}(c-2),upper middle) and vanish in $b<0$ (Fig.\ref{fig:main}(c-2),upper left). The $b$ dependence of the mirror Chern number $n_{{\cal M}_{(001)}}$ of this local model is calculated numerically and the result is shown in Fig.\ref{fig:main}(d-2). The mirror Chern number is changed at $b=0$, $n_{{\cal M}_{(001)}}=1$ for $b>0$, and $n_{{\cal M}_{(001)}}=0$ for $b<0$.

Next, we focus on the $k_y=0$ plane. The $+i$ mirror eigenvalue block is
\begin{equation}
    H_{\Sigma,+}^{(k_y=0)} = k_z \sigma_x + \chi k_x \sigma_y + (k_x^2 + k_z^2 + b) \sigma_z .
\end{equation}
The $y$ component of the Berry curvature of the occupied band is
\begin{equation}
    B_{\Sigma,+,y}^{(k_y=0)} = \frac{\chi}{2 R_{\Sigma,k_y=0}^3} (k_x^2 + k_z^2 - b) ,
\end{equation}
where the $R_{\Sigma,(k_y=0)}$ is defined as
\begin{equation}
    R_{\Sigma,k_y=0} = \sqrt{ k_z^2 + \chi^2 k_x^2 + (k_x^2 + k_z^2 + b)^2} .
\end{equation}
Contrary to the $k_x=0$ plane, the sharp peak feature of the Berry curvature on the $k_y=0$ plane is seen only when $b<0$. The mirror Chern number $n_{{\cal M}_{(110)}}$ of this local model also shows a contrasting behavior to $n_{{\cal M}_{(001)}}$ (Fig.\ref{fig:main}(d-2)), $n_{{\cal M}_{(110)}}=0$ for $b>0$, and $n_{{\cal M}_{(110)}}=1$ for $b<0$.

In both of the results on the $k_x=0$ plane and the $k_y=0$ plane, the nodal lines are considered as sources of the mirror Chern numbers. Furthermore, by comparing the two results, we can see that the $b=0$ is a phase transition point. Considering that the $b=0$ is the reconnection point of the nodal lines, the reconnection of nodal lines in the system without SOC corresponds to the topological phase transition in the system with SOC. Actually, when $b=0$, the model Eq.(\ref{eq:SlineSOC}) has a gapless point at $\bm{k}=0$. It is reasonable because generally a gap closing is required for a topological phase transition \cite{SCZ,TPT1}. 
Finally, we calculate the mirror Chern numbers of the whole BZ. The mirror numbers are $n_{{\cal M}_{(110)}}=0$ and  $n_{{\cal M}_{(001)}}=4$ for $b>0$, and  $n_{{\cal M}_{(110)}}=2$ and  $n_{{\cal M}_{(001)}}=0$ for $b<0$. Although the transition of the mirror Chern numbers is one by one in the local model, the transition in the whole BZ occurs between $n_{{\cal M}_{(001)}}=4$ and $n_{{\cal M}_{(110)}}=2$. It is because there are three symmetric planes for the (001) plane, but on the other hand, there are six symmetric planes for the (110) plane.

\textit{Discussion.}---
Although the local models are discussed separately, they are originated from a two-bands model in the whole BZ as we assumed first. In the first assumption, there are two different nodal line phases. One of them has nodal lines roughly located around the L-point (Fig.\ref{fig:main}(a-1)), and the other has nodal lines roughly located around the W-point (Fig.\ref{fig:main}(a-3)). The former phase includes the local model around the L-point and the local model around the $\Sigma$-line with $b<0$. The latter phase includes the local model around the W-point and the local model around the $\Sigma$-line with $b>0$. As shown in the local model calculations, when SOC is taken into account, the mirror Chern numbers are calculated by counting how many nodal rings locate on the mirror planes. The former nodal line semimetal phase is mapped to the topological crystalline insulator phase with $(n_{{\cal M}_{(110)}},n_{{\cal M}_{(001)}})=(2,0)$ (Fig.\ref{fig:main}(f-1)), and the latter nodal line semimetal phase is mapped to the topological crystalline insulator phase with $(n_{{\cal M}_{(110)}},n_{{\cal M}_{(001)}})=(0,4)$ (Fig.\ref{fig:main}(f-2)).
\begin{table}
    \centering
    \begin{tabular}{|c||c|c|c|}
        \multicolumn{4}{c}{Space group \#225 : $Fm\bar{3}m$} \\
        \hline
        $\mathbb{Z}_8$ & weak & $n_{{\cal M}_{(001)}}$ & $n_{{\cal M}_{(110)}}$ \\ \hline
        4 & 000 & 0 & 2 \\
        4 & 000 & 4 & 0 \\ \hline
    \end{tabular}
    \caption{Possible topological invariant combinations for $\mathbb{Z}_8=4$ phase in the space group \#225 with SOC \cite{CF1}.}
    \label{tab:fccind}
\end{table}
Now the obtained phases are topological crystalline insulator phases when SOC is taken into account, and it has been known that there are eight topological phases (including a trivial phase) indicated by the symmetry-based indicator $\mathbb{Z}_8$ in FCC (\#225) with SOC \cite{SBI}\footnote{Actually, four of them are prohibited by the compatibility relation in FCC.}. It is noteworthy how our obtained phases are classified by the symmetry-based indicator $\mathbb{Z}_8$. Actually, both of the obtained phases have the same indicator $\mathbb{Z}_8 = 4$ (Fig.\ref{fig:main}(e-1)). A previous study has given possible topological invariant combinations for each indicator \cite{CF1}, and that for $\mathbb{Z}_8 = 4$ is listed in Table \ref{tab:fccind}. It includes our obtained phases, of course. The reason why our obtained phases have the same indicator is that the gap closing on the phase transition point, or the reconnection of nodal lines in the system without SOC, occurs between the two bands with the same irreps. Since the symmetry-base indicator checks the irreps of occupied bands, it must be identical before and after the phase transition. However, our result has shown that when SOC is turned off, there is a clear difference between nodal lines penetrating the (110) mirror plane and nodal lines panatrating the (001) mirror plane. Additionally, considering that the nodal line is crossing the Q-line, it must penetrate one of the mirror planes. This means that when SOC is taken into account, the system cannot be topologically trivial and there are only two possible mirror Chern number combinations. 

\textit{Material examples.}---
In this section, we show the mapping proved in the previous sections is actually observed in material examples, FCC Ca and Ba. We calculate their band dispersions by the first-principles calculation for the both cases with and without SOC. These calculations are performed by Quantum ESPRESSO \cite{qe}, which uses the density functional theory \cite{dft1,dft2}. For the exchange-correlation term, generalized gradient approximation with Perdew–Burke–Ernzerhof parametrization \cite{PBE} is used. The plane wave energy cut-off is set to 50 Ry and the k-point grid on the BZ is taken as 8$\times$8$\times$8 mesh. The lattice constants are 5.601 \AA~for Ca and 6.0520 \AA~for Ba \cite{TopoMatDatabase,zhang2019catalogue}. With the output of the first-principles calculations, we calculate Wannier centers on the mirror invariant planes by using Z2pack code \cite{z2pack1,z2pack2}. Seeing the Wannier center flows, we calculate the mirror Chern numbers for the cases with SOC.

\begin{figure}
    \centering
    \includegraphics[width=8.5cm]{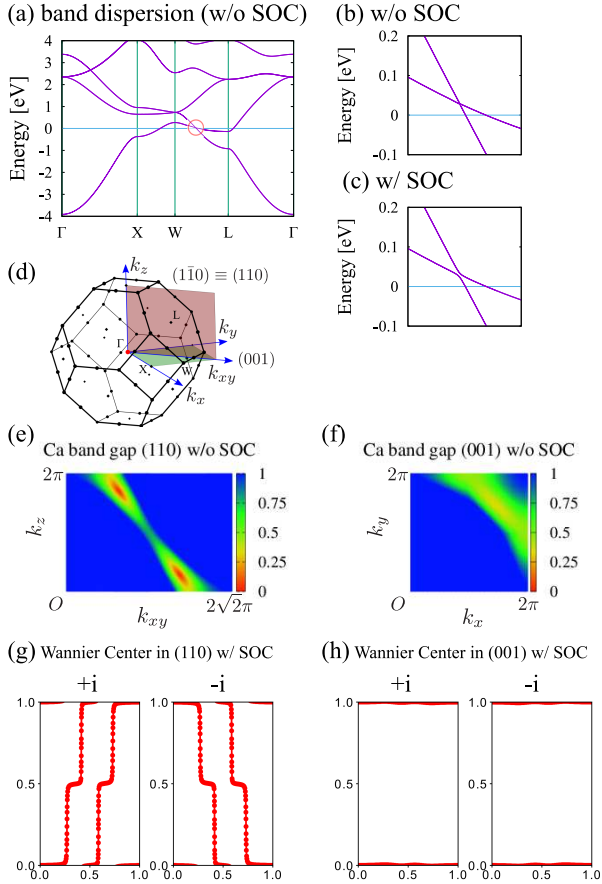}
    \caption{Band dispersion and Wannier centers of FCC Ca. (a) band dispersion (w/o SOC). A band crossing exists on L-W line and it is denoted with a red circle. (b) Enlarged band dispersion (w/o SOC) around the band crossing. (c) Enlarged band dispersion (w/ SOC). The two bands are slightly gaped out by the effect of SOC. (d) BZ and mirror invariant planes. (e) Band gap in the (110) mirror invariant plane (w/o SOC). The rectangle area is equivalent with the red rectangle in (d). There are two gapless point (red colored points) in this plane. (f) Band gap in the (001) mirror invariant plane (w/o SOC). The square area is equivalent with the green square in (d). There is no gapless point in this plane. (g) Wannier center in (110) mirror invariant plane (w/ SOC). The left (right) panel shows the Wannier center of the $+i$ ($-i$) mirror eigenvalue sector. The Wannier center is winding two times. (h) Wannier center in (001) mirror invariant plane (w/ SOC). The left (right) panel shows the Wannier center of the $+i$ ($-i$) mirror eigenvalue sector.}
    \label{fig:Ca}
\end{figure}

The result for Ca is shown in Fig.\ref{fig:Ca}. In Fig.\ref{fig:Ca}(a), which is a band dispersion without SOC, we can see a band crossing on the W-L line (denoted by a red circle). As explained before, this band crossing is a part of a nodal line. By checking the irreps of the occupied bands, it is confirmed that Ca without SOC is a nodal line semimetal with $\mathbb{Z}_4=0$. For the (110) and (001) mirror planes, the band gaps between the two bands which make the nodal line are shown in Fig.\ref{fig:Ca}(e) and (f). The rectangle area of Fig.\ref{fig:Ca}(e) (the square area of Fig.\ref{fig:Ca}(f)) is equivalent to the red rectangle (the green square) in Fig.\ref{fig:Ca}(d). In Ca, there are gapless points, which are plotted with red points, only on the (110) mirror plane. Therefore, Ca has Fig.\ref{fig:main}(a-1) type nodal lines when SOC is neglected. Let us introduce SOC into this system. The system is confirmed to have $\mathbb{Z}_8=4$. The band crossing on the W-L line slightly gaps out by the effect of SOC (Fig.\ref{fig:Ca}(c)). As a result, the band around -4 eV $\sim$ 0 eV has no gapless point and we can calculate the mirror Chern numbers on the band. The mirror Chern numbers are calculated as winding numbers of Wannier centers on the mirror invariant planes (Fig.\ref{fig:Ca}(g)(h)). We can see the winding number on the (110) mirror plane is 2, while that on the (001) mirror plane is 0. Therefore, Ca with SOC is a topological crystalline insulator with $(n_{{\cal M}_{(110)}},n_{{\cal M}_{(001)}})=(2,0)$ (Fig.\ref{fig:main}(f-1) phase).

\begin{figure}
    \centering
    \includegraphics[width=8.5cm]{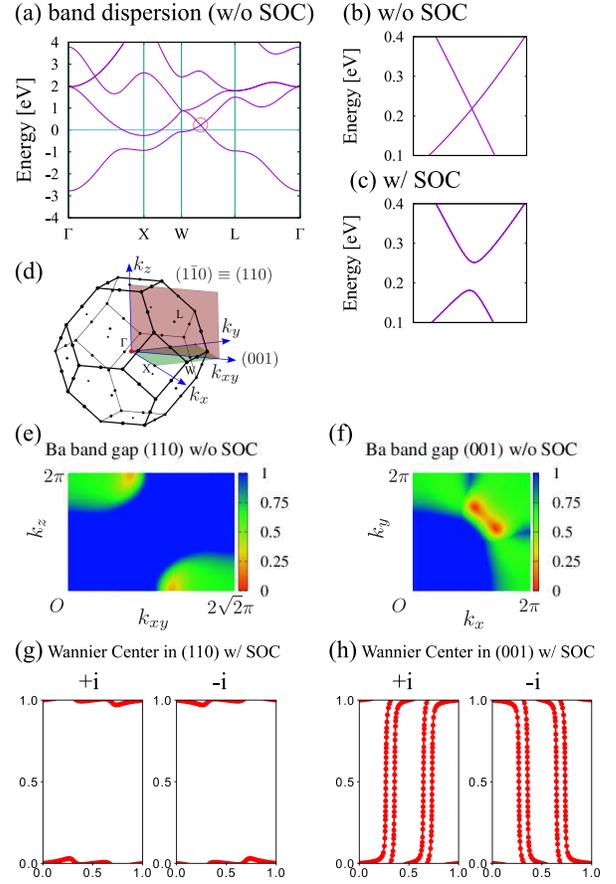}
    \caption{Band dispersion and Wannier centers of FCC Ba. (a) band dispersion (w/o SOC). A band crossing exists on L-W line and it is denoted with a red circle. (b) Enlarged band dispersion (w/o SOC) around the band crossing. (c) Enlarged band dispersion (w/ SOC). The two bands are gaped out by the effect of SOC. (d) BZ and mirror invariant planes. (e) Band gap in the (110) mirror invariant plane (w/o SOC). The rectangle area is equivalent with the red rectangle in (d). There is no gapless point in this plane. (f) Band gap in the (001) mirror invariant plane (w/o SOC). The square area is equivalent with the green square in (d). There are two gapless point (red colored points) in this plane. (g) Wannier center in (110) mirror invariant plane (w/ SOC). The left (right) panel shows the Wannier center of the $+i$ ($-i$) mirror eigenvalue sector. (h) Wannier center in (001) mirror invariant plane (w/ SOC). The left (right) panel shows the Wannier center of the $+i$ ($-i$) mirror eigenvalue sector. The Wannier center is winding four times.}
    \label{fig:Ba}
\end{figure}

Next, we move to the result for Ba shown in Fig.\ref{fig:Ba}. In Fig.\ref{fig:Ba}(a), we can see a band crossing on the W-L line also in Ba (denoted by a red circle). This band crossing is also a part of a nodal line and Ba is also confirmed to be a nodal semimetal with $\mathbb{Z}_4=0$. The band gaps between the two bands on the (110) and (001) mirror planes are shown in Fig.\ref{fig:Ba}(e)(f). In contrast to the Ca, there are gapless points only on the (001) mirror plane in Ba. Therefore, Ba has Fig.\ref{fig:main}(a-3) type nodal lines when SOC is neglected. Let us introduce SOC into this system. The system is confirmed to have $\mathbb{Z}_8=4$. The band crossing on the W-L line gaps out by the effect of SOC (Fig.\ref{fig:Ba}(c)). As a result, the band around -3 eV $\sim$ 0 eV has no gapless point. The calculated Wannier centers on the mirror invariant planes are shown in Fig.\ref{fig:Ba}(g)(h). We can see the winding number on the (001) mirror plane is 4, while that on the (110) mirror plane is 0. Therefore, Ba with SOC is a topological crystalline insulator with $(n_{{\cal M}_{(110)}},n_{{\cal M}_{(001)}})=(0,4)$ (Fig.\ref{fig:main}(f-2) phase).

\textit{Conclusion.}---
In this paper, we discussed what kind of topological phases emerge from nodal line semimetals with $\mathbb{Z}_4=0$ in FCC when SOC is introduced. Our model calculation proved the $\mathbb{Z}_4=0$ nodal line semimetals are mapped to topological crystalline insulator phases. Furthermore, the difference in nodal line configurations corresponds to the difference in topological invariants. We also showed that this mapping is actually observed in some material examples by using the first-principles calculations. In this mapping, two topological crystalline insulator phases which can not be distinguished by previous diagnostic methods are divided. Although our calculation is still a case study in FCC, this fact can lead to a subdividing classification theory.

\textit{Note added 1.}---
Our model with nodal lines in this study has $\mathbb{Z}_4=0$ when SOC is neglected. On the other hand, when SOC is taken into account, $\mathbb{Z}_8=4$. Considering that "true trivial phase" in \#225 should have $\mathbb{Z}_4=0$ (w/o SOC) and $\mathbb{Z}_8=0$ (w/ SOC), this fact suggests that there is "another $\mathbb{Z}_2$ index" and it will distinguish $\mathbb{Z}_4=0$ (w/o SOC) with and without nodal lines. However, this point remains as a future work.

\textit{Note added 2.}---
As a parallel work, the mapping rule is studied in the body-centered tetragonal system (the space group $\#$139 ($I4/mmm$)) \cite{ITparallelBCT}. In this parallel work, a similar mapping from nodal lines to mirror Chern numbers is reported.

\textit{Acknowledgement.}---
We acknowledge the many fruitful discussions with Nobuyuki Okuma, Hiroyasu Matsuura and Masao Ogata. I.T. was supported by KAKENHI 17H02912 from JSPS and by the Japan Society for the Promotion of Science through the Program for Leading Graduate Schools (MERIT).

\bibliography{reference}
\bibliographystyle{apsrev4-2}

\clearpage
\appendix
\section{\label{SMsec:PG}Point groups and irreducible representations}
\begin{figure}
    \centering
    \includegraphics[width=5cm]{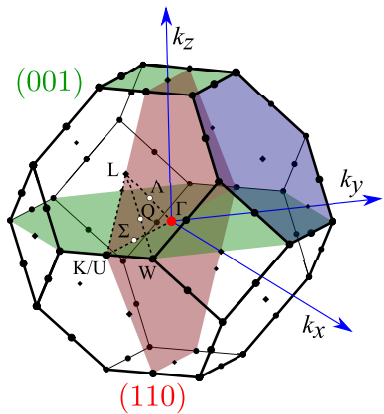}
    \caption{Brillouin zone of face-centered cubic lattice (Space group \#225, $Fm\bar{3}m$). The (001) mirror plane and the (110) mirror plane is shown as green and red planes, respectively. The hexagonal surface of the BZ is shown as a blue plane for convenience, though it is not a high-symmetry plane.}
    \label{fig:sfccbz}
\end{figure}

Here we describe the definition of the point groups and irreducible representations (irreps) we use in the main article \cite{BC}. The L-point, W-point, and $\Sigma$-line are the high-symmetry points (line) in the Brillouin zone (BZ) of the face-centered cubic (FCC) lattice (Fig.\ref{fig:sfccbz}).

\begin{table}[h]
    \centering
    \begin{tabular}{|c||c|c|c|c|c|c|}
        \multicolumn{7}{c}{L-point} \\ \hline
        D$_{\mathrm{3d}}$ & $E$ & 2$C_3$ & 3$C'_2$ & $I$ & 2$IC_3$ & 3 $m_v$ \\
        \hline \hline
        $\Gamma_1^+$ & 1 & 1 & 1 & 1 & 1 & 1 \\ \hline
        $\Gamma_2^+$ & 1 & 1 & -1 & 1 & 1 & -1 \\ \hline
        $\Gamma_3^+$ & 2 & -1 & 0 & 2 & -1 & 0 \\ \hline
        $\Gamma_1^-$ & 1 & 1 & 1 & -1 & -1 & -1 \\ \hline
        $\Gamma_2^-$ & 1 & 1 & -1 & -1 & -1 & 1 \\ \hline
        $\Gamma_3^-$ & 2 & -1 & 0 & -2 & 1 & 0 \\ \hline
    \end{tabular}
    \caption{Point group D$_{\mathrm{3d}}$ and its irreps. This point group is a little group of the L-point. The $m_v$ is the (110) mirror, and the $C'_2$ is the rotation around the Q-line.}
    \label{tab:PG-L}
\end{table}

\begin{table}[h]
    \centering
    \begin{tabular}{|c||c|c|c|c|c|}
        \multicolumn{6}{c}{W-point} \\ \hline
        D$_{\mathrm{2d}}$ & $E$ & 2$IC_4$ & $C_2$ & 2$C'_2$ & 2 $m_d$ \\
        \hline \hline
        $\Gamma_1$ & 1 & 1 & 1 & 1 & 1 \\ \hline
        $\Gamma_2$ & 1 & 1 & 1 & -1 & -1 \\ \hline
        $\Gamma_3$ & 1 & -1 & 1 & 1 & -1 \\ \hline
        $\Gamma_4$ & 1 & -1 & 1 & -1 & 1 \\ \hline
        $\Gamma_5$ & 2 & 0 & -2 & 0 & 0 \\ \hline
    \end{tabular}
    \caption{Point group D$_{\mathrm{2d}}$ and its irreps. This point group is a little group of the W-point. The $m_d$ is the (001) mirror, and the $C'_2$ is the Q-line.}
    \label{tab:PG-W}
\end{table}

\begin{table}[h]
    \centering
    \begin{tabular}{|c||c|c|c|c|}
        \multicolumn{5}{c}{$\Sigma$-line} \\ \hline
        C$_{\mathrm{2v}}$ & $E$ & $C_2$ & $m_y$ & $m_x$ \\
        \hline \hline
        $\Gamma_1$ & 1 & 1 & 1 & 1 \\ \hline
        $\Gamma_2$ & 1 & 1 & -1 & -1 \\ \hline
        $\Gamma_3$ & 1 & -1 & 1 & -1 \\ \hline
        $\Gamma_4$ & 1 & -1 & -1 & 1 \\ \hline
    \end{tabular}
    \caption{Point group C$_{\mathrm{2v}}$ and its irreps. This point group is a little group of the $\Sigma$-line. The $m_y$ is the (110) mirror, and the $m_x$ is the (001) mirror.}
    \label{tab:PG-S}
\end{table}

\section{\label{SMsec:model}Construction of the local models}
Here we explain how we construct the two by two local models on the L-point, the W-point, and the $\Sigma$-line.
First of all, we consider nodal lines protected by time-reversal(TR) and inversion symmetries \cite{nodal,protnode}. To construct concrete local models, we briefly review how TR+inversion protected nodal lines emerge. Generally, TR+inversion symmetry prohibits the existence of a point node in the momentum space. Therefore, when the band dispersion has a degenerate point on a high-symmetry line, it must be a part of a nodal line. 

In our target system, which is based on some realistic materials \cite{NLHirayama,ITSnSe}, the nodal line emerges from a degenerated point on the $C_2$ rotation invariant line (the Q-line).
\subsection{\label{SMsec:modelirrep}Irreps of the each band}
\subsubsection{L-point}
The two bands must satisfy the following conditions.
\begin{itemize}
    \item The two bands have the same mirror eigenvalue for $m_v$ (no nodal line on the mirror plane).
    \item The two bands have different $C'_2$ rotation eigenvalues.
    \item The two bands are non-degenerate bands.
\end{itemize}
The pair of two bands which satisfy these conditions is $\{ \Gamma_1^+ , \Gamma_2^- \}$ or $\{ \Gamma_2^+ , \Gamma_1^- \}$

\subsubsection{W-point}
The two bands must satisfy the following conditions.
\begin{itemize}
    \item The two bands have the same mirror eigenvalue for $m_d$ (no nodal line on the mirror plane).
    \item The two bands have different $C'_2$ rotation eigenvalues.
    \item The two bands are non-degenerate bands.
\end{itemize}
The pair of two bands which satisfy these conditions is $\{ \Gamma_1 , \Gamma_4 \}$ or $\{ \Gamma_2 , \Gamma_3 \}$

\subsubsection{$\Sigma$-line}
The two bands must satisfy the following conditions.
\begin{itemize}
    \item The two bands have the same mirror eigenvalue for $m_x$ (no nodal line on the mirror plane).
    \item The two bands have the same mirror eigenvalue for $m_y$ (no nodal line on the mirror plane).
    \item The two bands are non-degenerate bands.
\end{itemize}
The pair of two bands which satisfy these conditions is $\{ \Gamma_1 , \Gamma_1 \}$ or $\{ \Gamma_2 , \Gamma_2 \}$ or $\{ \Gamma_3 , \Gamma_3 \}$ or $\{ \Gamma_4 , \Gamma_4 \}$.

\subsection{\label{SMsec:modelconn}Band connection}
\begin{figure}
    \centering
    \includegraphics[width=6cm]{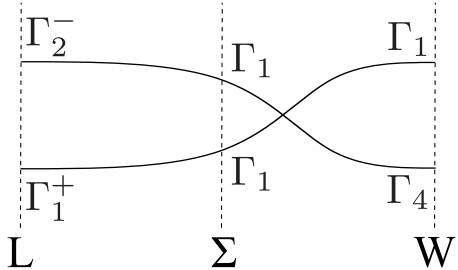}
    \caption{Schematic picture of the band dispersion of the case (1) in Table \ref{tab:ireepsset}.}
    \label{fig:SMband}
\end{figure}

By considering the compatibility relation, we can decide how the bands in each high-symmetry points (line) are connected. When the system has nodal lines, the irreps of the occupied band on the L-point and the W-point must violate the compatibility relation along the L-W line, i. e. the eigenvalues of the $C_2$ rotation along the L-W are different in both ends. The possible combinations are given in Table \ref{tab:ireepsset}. A schematic picture of the band dispersion of the case (1) is shown in Fig.\ref{fig:SMband}.
\begin{table}
    \centering
    \begin{tabular}{|c||c|c|c|}
        \hline
         & L-point & $\Sigma$-line & W-point \\
        \hline
        (1) & $\{ \Gamma_1^+ , \Gamma_2^- \}$ & $\{ \Gamma_1 , \Gamma_1 \}$ & $\{ \Gamma_4 , \Gamma_1 \}$ \\ \hline
        (2) & $\{ \Gamma_1^+ , \Gamma_2^- \}$ & $\{ \Gamma_3 , \Gamma_3 \}$ & $\{ \Gamma_3 , \Gamma_2 \}$ \\ \hline
        (3) & $\{ \Gamma_2^+ , \Gamma_1^- \}$ & $\{ \Gamma_2 , \Gamma_2 \}$ & $\{ \Gamma_3 , \Gamma_2 \}$ \\ \hline
        (4) & $\{ \Gamma_2^+ , \Gamma_1^- \}$ & $\{ \Gamma_4 , \Gamma_4 \}$ & $\{ \Gamma_4 , \Gamma_1 \}$ \\ \hline
    \end{tabular}
    \caption{Possible set of irreps of bands on each high-symmetry points.}
    \label{tab:ireepsset}
\end{table}

\subsection{\label{SMsec:modelder}Derivation of the local model}
Here we explain how we derive the local models on each high-symmetry point (line). For the set of irreps, we assume the case (1) in Table \ref{tab:ireepsset}. However, it is proved later that the choice of the set does not matter in the construction of the local models.
Now we construct a two-bands model, and thus we derive the coefficients of the Pauli matrices to satisfy the symmetric restriction of the system. We consider the $\bm{k} \cdot \bm{p}$-perturbation within the second-order of $k$.

\subsubsection{L-point}
\begin{figure}
    \centering
    \includegraphics[width=7cm]{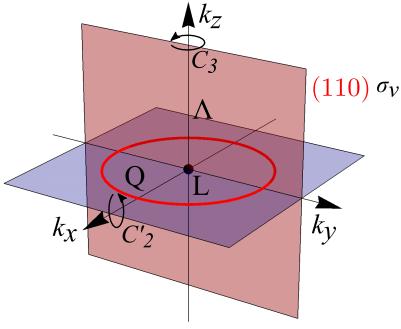}
    \caption{Coordinate system of the local model around the L-point.}
    \label{fig:Lcood}
\end{figure}

The coordinate system of the local model around the L-point is shown in Fig.\ref{fig:Lcood}. The origin is placed at the L-point and the $k_z$ axis is parallel to the $\Lambda$-line, which is the $C_3$ rotation axis. The $k_z$ axis is perpendicular to the hexagonal face of the BZ (blue plane). The $k_x$ axis is parallel to the Q-line, which is the $C'_2$ rotation axis. The $m_v$ mirror invariant plane, which is the (110) plane, is the $k_x=0$ plane. In Table \ref{tab:Lterm}, we show how each term is transformed by the symmetry operators of the little group of the L-point. When we consider a model within the second-order of $k$, the $C_3$ rotation symmetry requires the model to be isotropic in the $k_x$-$k_y$ plane. It is assumed that the two bases have different eigenvalues for $C'_2$, while they have the same eigenvalues for $m_v$. Therefore, the operators are written as $\sigma_z$ for $C'_2$ and $I$, and $\sigma_0$ for $m_v$. The local model around the L-point is written as
\begin{equation}
    H(\bm{k}) = k_z \sigma_x + (k_x^2 + k_y^2 + k_z^2 + \mathrm{const.}) \sigma_z .
\end{equation}
We neglect some degree of freedom to make the model simple, e. g., a constant energy shift by $\sigma_0$, and the coefficient of the $k_z$ term. It is because they are negligible in our calculation of the topological invariants.

The local model with SOC (amplitude $\chi$) is written as
\begin{equation}
    \begin{split}
    H_{\mathrm{soc}}(\bm{k}) =& k_z \sigma_x s_0 + (k_x^2 + k_y^2 + k_z^2 + \mathrm{const.}) \sigma_z s_0 \\
    & + \chi \sigma_y (-k_y s_x + k_x s_y) .
    \end{split}
\end{equation}
The operator of the mirror $m_v$ is given as $i\sigma_0 s_x$, and the operator of $C'_2$ is given as $i\sigma_z s_x$. 

\begin{table}
    \centering
    \begin{tabular}{|c||c|c|c|c|c|c|}
        \multicolumn{7}{c}{L-point} \\ \hline
         & $E$ & $2 C_3$ & $3 C'_2$ & $I$ & $2IC_3$ & $3 m_v$ \\ \hline \hline
        operator & $\sigma_0$ & & $\sigma_z$ & $\sigma_z$ & & $\sigma_0$ \\ \hline \hline
        1 (const.) & 1 & 1 & 1 & 1 & 1 & 1 \\ \hline \hline
        $k_x$ & $k_x$ & * & $k_x$ & $-k_x$ & * & $-k_x$ \\ \hline
        $k_y$ & $k_y$ & * & $-k_y$ & $-k_y$ & * & $k_y$ \\ \hline
        $k_z$ & $k_z$ & $k_z$ & $-k_z$ & $-k_z$ & $-k_z$ & $k_z$ \\ \hline \hline
        $k_x^2+k_y^2$ & $k_x^2+k_y^2$ & $k_x^2+k_y^2$ & $k_x^2+k_y^2$ & $k_x^2+k_y^2$ & $k_x^2+k_y^2$ & $k_x^2+k_y^2$ \\ \hline
        $k_z^2$ & $k_z^2$ & $k_z^2$ & $k_z^2$ & $k_z^2$ & $k_z^2$ & $k_z^2$ \\ \hline 
        \multicolumn{7}{r}{* $k_x$ and $k_y$ are prohibited by the $C_3$}
    \end{tabular}
    \caption{Transformation of each term by the symmetric operator of the little group of the L-point. $k_x$, $k_y$, $k_x k_y$, $k_y k_z$, and $k_z k_x$ are prohibited by the $C_3$}
    \label{tab:Lterm}
\end{table}

\subsubsection{W-point}
\begin{figure}
    \centering
    \includegraphics[width=7cm]{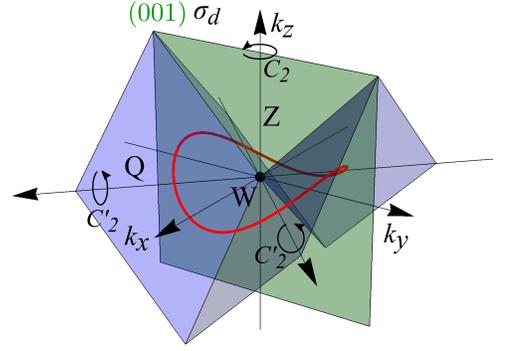}
    \caption{Coordinate system of the local model around the W-point.}
    \label{fig:Wcood}
\end{figure}

The coordinate system of the local model around the W-point is shown in Fig.\ref{fig:Wcood}. The origin is placed at the W-point and the $k_z$ axis is parallel to the Z-line, which is the $C_2$ rotation axis. The $m_d$ mirror invariant plane, which is the (001) plane, is the $k_x=0$ plane. The Q-line, which is $C'_2$ rotation axis, is a line represented as $\{ k_x + k_y =0 ~\mathrm{and}~ k_z=0\}$. In Table \ref{tab:Lterm}, we show how each term is transformed by the symmetry operators of the little group of the W-point. As explained before in the derivation around the L-point, the operator of $m_d$ and $C'_2$ are assumed as $\sigma_0$ and $\sigma_z$, respectively. The local model around the W-point is written as
\begin{equation}
    \begin{split}
    H(\bm{k}) =& (k_z + a k_x^2 - a k_y^2) \sigma_x  \\
    & + (k_x^2 + k_y^2 + k_z^2 + \mathrm{const.}) \sigma_z ,
    \end{split}
\end{equation}
where $a$ is a real constant.

The local model with SOC (amplitude $\chi$) is written as
\begin{equation}
    \begin{split}
    H_{\mathrm{soc}}(\bm{k}) =& (k_z + a k_x^2 - a k_y^2) \sigma_x s_0 \\
    & + (k_x^2 + k_y^2 + k_z^2 + \mathrm{const.}) \sigma_z s_0 \\
    & + \chi \sigma_y (-k_y s_x + k_x s_y) .
    \end{split}
\end{equation}
The operator of the mirror $m_d$ is given as $i \sigma_0 s_x$, and the operator of $C'_2$ is given as $i\sigma_z \frac{1}{\sqrt{2}} (s_x + s_y)$.

\begin{table}
    \centering
    \begin{tabular}{|c||c|c|c|c|c|}
        \multicolumn{6}{c}{W-point} \\ \hline
         & $E$ & $I C_4$ & $C_2$ & $2 C'_2$ & $2m_d$ \\ \hline \hline
        operator & $\sigma_0$ & & & $\sigma_z$ & $\sigma_0$ \\ \hline \hline
        1 (const.) & 1 & 1 & 1 & 1 & 1 \\ \hline \hline
        $k_x$ & $k_x$ & $-k_y$ & $-k_x$ & $k_y$ & $-k_x$ \\ \hline
        $k_y$ & $k_y$ & $k_x$ & $-k_y$ & $k_x$ & $k_y$ \\ \hline
        $k_z$ & $k_z$ & $-k_z$ & $k_z$ & $-k_z$ & $k_z$ \\ \hline \hline
        $k_x^2$ & $k_x^2$ & $k_y^2$ & $k_x^2$ & $k_y^2$ & $k_x^2$ \\ \hline
        $k_y^2$ & $k_y^2$ & $k_x^2$ & $k_y^2$ & $k_x^2$ & $k_y^2$ \\ \hline
        $k_z^2$ & $k_z^2$ & $k_z^2$ & $k_z^2$ & $k_z^2$ & $k_z^2$ \\ \hline \hline
        $k_x k_y$ & $k_x k_y$ &$-k_x k_y$ & $k_x k_y$ & $k_x k_y$ & $-k_x k_y$ \\ \hline
        $k_y k_z$ & $k_y k_z$ & $-k_x k_z$ & $-k_y k_z$ & $-k_x k_z$ & $k_y k_z$ \\ \hline
        $k_z k_x$ & $k_z k_x$ & $k_y k_z$ & $-k_z k_x$ & $-k_y k_z$ & $-k_z k_x$ \\ \hline
    \end{tabular}
    \caption{Transformation of each term by the symmetric operator of the little group of the W-point.}
    \label{tab:Wterm}
\end{table}

\subsubsection{$\Sigma$-line}
\begin{figure}
    \centering
    \includegraphics[width=7cm]{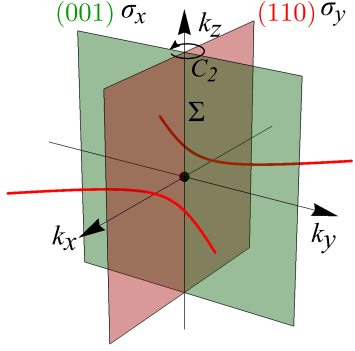}
    \caption{Coordinate system of the local model around the $\Sigma$-line.}
    \label{fig:Sigcood}
\end{figure}

The coordinate system of the local model around the $\Sigma$-line is shown in Fig.\ref{fig:Sigcood}. Now the local model should be constructed to represent the reconnection of nodal lines by tuning a parameter. The symmetry of the FCC lattice requires that the reconnection of the nodal lines should occur on the $\Sigma$-line. The origin is placed at the point where the reconnection of nodal lines occurs. The $k_z$ axis is parallel to the $\Sigma$-line. The (001) mirror plane and the (110) mirror plane are the $k_x=0$ plane and the $k_y=0$ plane, respectively. In Table \ref{tab:Sigterm}, we show how each term is transformed by the symmetry operators of the little group of the $\Sigma$-line. Now both of the (001) mirror operator and the (110) mirror operator are assumed as $\sigma_0$. The local model around the $\Sigma$-line can be given with tunable parameter $b$ as
\begin{equation}
    H(\bm{k}) = k_z  \sigma_x + (k_x^2 - k_y^2 + k_z^2 + b) \sigma_z .
\end{equation}
In this model, hyperbolic nodal lines emerge and a reconnection of them occurs at the origin when $b=0$.

The local model with SOC (amplitude $\chi$) is written as
\begin{equation}
    \begin{split}
    H_{\mathrm{soc}}(\bm{k}) =& k_z \sigma_x s_0 + (k_x^2 - k_y^2 + k_z^2 + b) \sigma_z s_0 \\
    & + \chi \sigma_y (-k_y s_x + k_x s_y) .
    \end{split}
\end{equation}
The operator of the (001) mirror is given as $i \sigma_0 s_x$, and the operator of the (110) mirror is given as $i \sigma_0 s_y$.

\begin{table}
    \centering
    \begin{tabular}{|c||c|c|c|c|}
        \multicolumn{5}{c}{$\Sigma$-line} \\ \hline
         & $E$ & $C_2$ & $m_x$ & $m_y$ \\ \hline \hline
        operator & $\sigma_0$ & & $\sigma_0$ & $\sigma_0$ \\ \hline \hline
        1 (const.) & 1 & 1 & 1 & 1 \\ \hline \hline
        $k_x$ & $k_x$ & $-k_x$ & $-k_x$ & $k_x$ \\ \hline
        $k_y$ & $k_y$ & $-k_y$ & $k_y$ & $-k_y$ \\ \hline
        $k_z$ & $k_z$ & $k_z$ & $k_z$ & $k_z$ \\ \hline \hline
        $k_x^2$ & $k_x^2$ & $k_x^2$ & $k_x^2$ & $k_x^2$ \\ \hline
        $k_y^2$ & $k_y^2$ & $k_y^2$ & $k_y^2$ & $k_y^2$ \\ \hline
        $k_z^2$ & $k_z^2$ & $k_z^2$ & $k_z^2$ & $k_z^2$ \\ \hline \hline
        $k_x k_y$ & $k_x k_y$ &$k_x k_y$ & $-k_x k_y$ & $-k_x k_y$ \\ \hline
        $k_y k_z$ & $k_y k_z$ & $-k_y k_z$ & $k_y k_z$ & $-k_y k_z$ \\ \hline
        $k_z k_x$ & $k_z k_x$ & $-k_z k_x$ & $-k_z k_x$ & $k_z k_x$ \\ \hline
    \end{tabular}
    \caption{Transformation of each term by the symmetric operator of the little group of the $\Sigma$-line.}
    \label{tab:Sigterm}
\end{table}

\subsubsection{About other sets of irreps}
We considered the (1) case in Table \ref{tab:ireepsset} above. When we choose another case, we need to replace some mirror operator with $-\sigma_0$, instead of $\sigma_0$. However, it is easy to see that this replacement makes no difference in the derived model. Therefore, we can say that the derived model is describing all cases in Table \ref{tab:ireepsset}.

\section{\label{SMsec:MCN}Calculation of the mirror Chern number}
Here we show the detail of the calculation of the Berry curvature and the mirror Chern number. 

\subsection{L-point}
The local model with SOC is written as
\begin{equation}
    \begin{split}
    H_{L,\mathrm{soc}} =& k_z \sigma_x s_0 + (k_x^2 + k_y^2 + k_z^2 - \Delta^2) \sigma_z s_0 \\
    & + \chi \sigma_y (-k_y s_x + k_x s_y),
    \end{split}
\end{equation}
where $\Delta$ is a real positive constant. When SOC is neglected ($\chi=0$), the gapless points are given as a solution of
\begin{equation}
    \left\{
    \begin{array}{l}
        k_z=0 \\
        k_x^2 + k_y^2 + k_z^2 - \Delta^2 = 0
    \end{array}
    \right. .
\end{equation}
The solution of these equations is a ring with a radius $\Delta$ on the $k_x$-$k_y$ plane.

\begin{figure*}
    \centering
    \includegraphics[width=14cm]{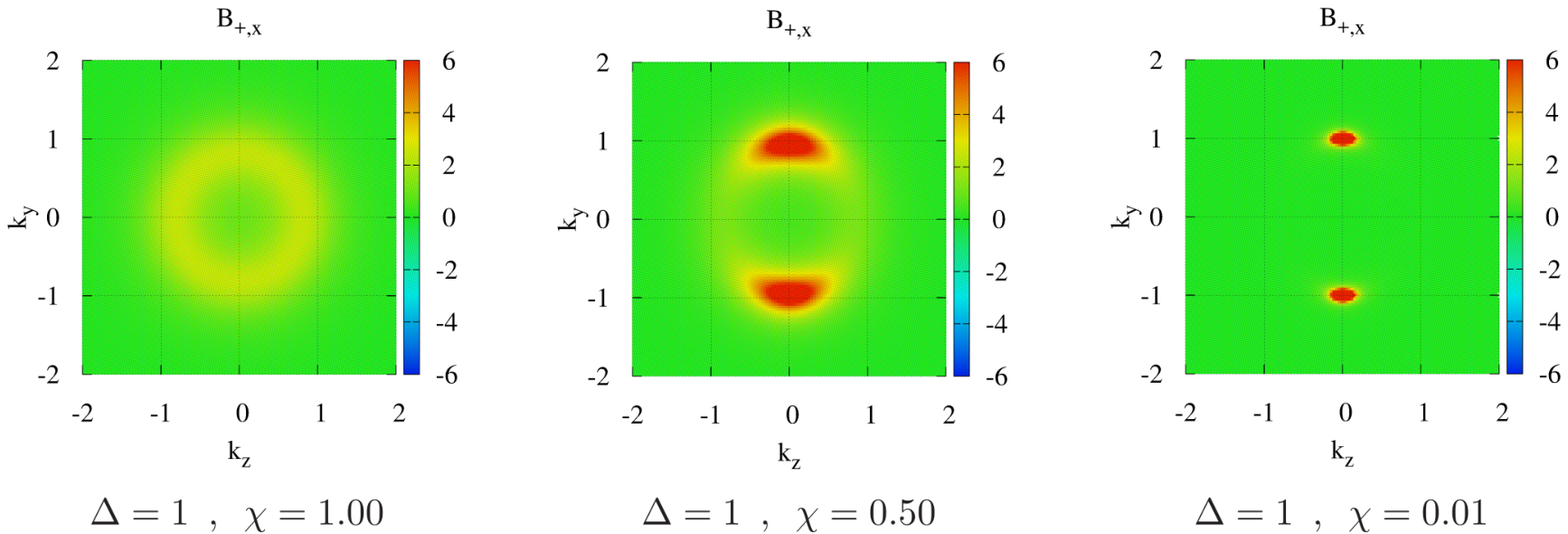}
    \caption{$\chi$ dependence of $B_{L,+,x}$.}
    \label{fig:Lberry}
\end{figure*}

On the (110) plane (the $k_x=0$ plane), the model is rewritten as
\begin{equation}
    \begin{split}
    H_{L,\mathrm{soc}} =& k_z \sigma_x s_0 + (k_y^2 + k_z^2 - \Delta^2) \sigma_z s_0 \\
    & - \chi \sigma_y k_y s_x .
    \end{split}
\end{equation}
By using a unitary transformation $U=\sigma_0 \frac{1}{\sqrt{2}} (s_x + s_z)$, the model is block diagonalized as
\begin{equation}
    \begin{split}
    U^\dagger H_{L,\mathrm{soc}} U =& k_z \sigma_x s_0 + (k_y^2 + k_z^2 - \Delta^2) \sigma_z s_0 \\
    & - \chi \sigma_y k_y s_z .
    \end{split}
\end{equation}
The (110) mirror operator is also transformed as $U^\dagger i \sigma_0 s_x U = i \sigma_0 s_z$. Now the two blocks are corresponding to the blocks with $+i$ and $-i$ mirror eigenvalues, respectively. They are explicitly given as
\begin{equation}
    H_{L,+} = X_L \sigma_x - Y_L \sigma_y + Z_L \sigma_z ,
\end{equation}
\begin{equation}
    H_{L,-} = X_L \sigma_x + Y_L \sigma_y + Z_L \sigma_z ,
\end{equation}
\begin{equation}
    X_L = k_z ~~,~~ Y_L = \chi k_y ~~,~~ Z_L = k_y^2 + k_z^2 - \Delta^2 .
\end{equation}
The Berry connection of the occupied band of $H_{L,+}$ is calculated as
\begin{equation}
    \begin{split}
        A_{L,+,y} = \frac{-\chi k_z}{2R_L (R_L - Z_L)} ,\\
        A_{L,+,z} = \frac{\chi k_y}{2R_L (R_L - Z_L)} ,
    \end{split}
\end{equation}
where $R_L$ is defined as
\begin{equation}
    R_L = \sqrt{ X_L^2 + Y_L^2 + Z_L^2} .
\end{equation}
The $x$ component of the Berry curvature is given as
\begin{equation}
    \begin{split}
    B_{L,+,x} =& \frac{\partial A_{L,+,z}}{\partial k_y} - \frac{\partial A_{L,+,y}}{\partial k_z} \\
    =& \frac{\chi}{2R_L^3} (Z_L + 2 \Delta^2) .
    \end{split}
\end{equation}
The $\chi$ dependence of $B_{L,+,x}$ is shown in Fig.\ref{fig:Lberry}. When $\chi$ is large, for example $\chi=1.00$, $B_{L,+,x}$ is spread widely. As $\chi$ getting smaller, sharper peaks appear on the points where the nodal line penetrates when SOC is neglected. The Chern number of the occupied band of the $+i$ block, $C_{L,+}$, is calculated by integrating the Berry connection on a circle path $(k_y,k_z)=(k\cos \theta, k\sin \theta)$,
\begin{equation}
    \begin{split}
    C_{L,+} =& \frac{1}{2\pi} \oint d \bm{k} \cdot \bm{A} \\
    =& \frac{1}{2\pi} \int_0^{2\pi} d \theta \frac{\chi k^2}{k^2 (\cos^2 \theta + \chi^2 \sin^2 \theta)} \\
    =& \frac{\chi}{| \chi |} .
    \end{split}
\end{equation}
Here we assume $k \gg 1$, and then $R_L \sim |Z_L|$ and $R_L - Z_L \sim \frac{X_L^2 + Y_L^2}{2|Z_L|}$.
Generally, it is known that the Chern number of the $-i$ block satisfies $C_{L,-} = -C_{L,+}$. Therefore, the mirror Chern number $n_{{\cal M}_{(110)}}$ of this local model is given as
\begin{equation}
    \begin{split}
    n_{L,{\cal M}_{(110)}} &= \frac{C_{L,+}-C_{L,-}}{2} \\
    &= \frac{\chi}{| \chi |} = \left\{
    \begin{array}{ll}
        1 & (\chi>0) \\
        -1 & (\chi<0)
    \end{array} \right. .
    \end{split}
\end{equation}

\subsection{W-point}
The local model with SOC is written as
\begin{equation}
    \begin{split}
    H_{W,\mathrm{soc}} =& (k_z + a k_x^2 - a k_y^2) \sigma_x s_0 \\
    & + (k_x^2 + k_y^2 + k_z^2 - \Delta^2) \sigma_z s_0 \\
    & + \chi \sigma_y (-k_y s_x + k_x s_y) ,
    \end{split}
\end{equation}
where $\Delta$ and $a$ are real positive constants. When SOC is neglected ($\chi=0$), the gapless points are given as a solution of
\begin{equation}
    \left\{
        \begin{array}{l}
            k_z + a k_x^2 - a k_y^2 = 0 \\
            k_x^2 + k_y^2 + k_z^2 - \Delta^2 = 0
        \end{array} \right. .
\end{equation}
The solution of these equations is a "oscillating ring" around the origin.

\begin{figure*}
    \centering
    \includegraphics[width=14cm]{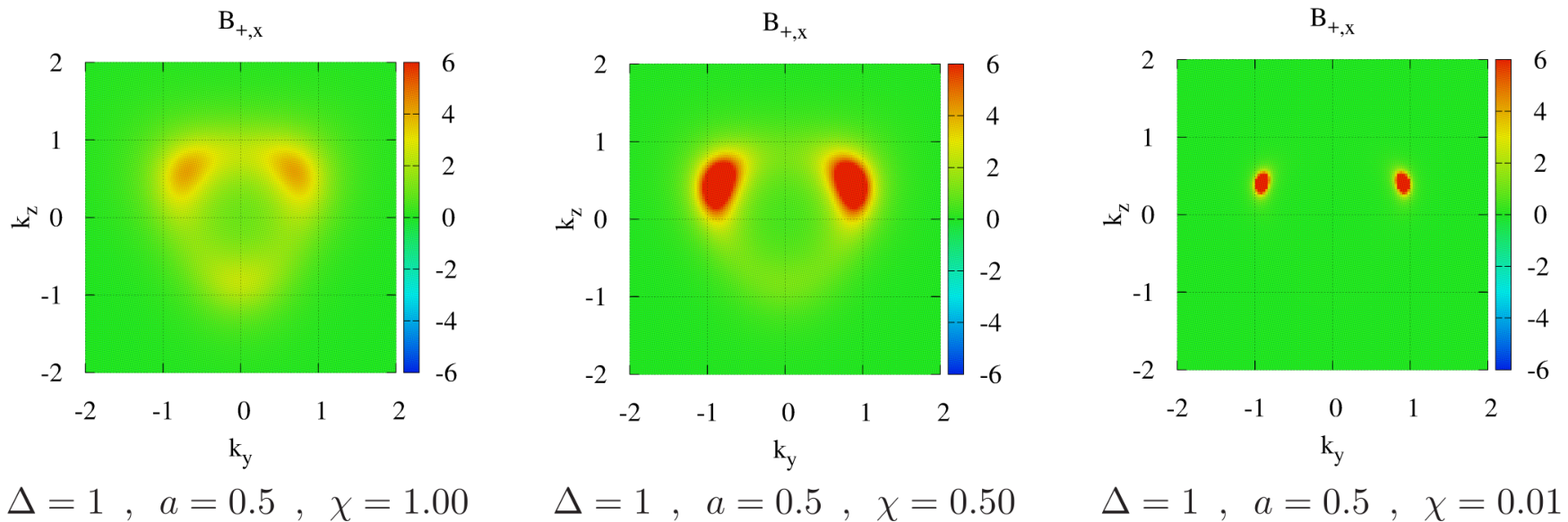}
    \caption{$\chi$ dependence of $B_{W,+,x}$.}
    \label{fig:Wberry}
\end{figure*}

On the (001) plane (the $k_x=0$ plane), the model is rewritten as
\begin{equation}
    \begin{split}
    H_{W,\mathrm{soc}} =& (k_z - a k_y^2) \sigma_x s_0 +  (k_y^2 + k_z^2 - \Delta^2) \sigma_z s_0 \\
    & - \chi \sigma_y k_y s_x .
    \end{split}
\end{equation}
By using a unitary transformation $U=\sigma_0 \frac{1}{\sqrt{2}} (s_x + s_z)$, the model is block diagonalized as
\begin{equation}
    \begin{split}
    U^\dagger H_{W,\mathrm{soc}} U =& (k_z - a k_y^2) \sigma_x s_0 +  (k_y^2 + k_z^2 - \Delta^2) \sigma_z s_0 \\
    & - \chi \sigma_y k_y s_z .
    \end{split}
\end{equation}
The (001) mirror operator is also transformed as $U^\dagger i \sigma_0 s_x U = i \sigma_0 s_z$. The two block with $+i$ and $-i$ mirror eigenvalues are explicitly given as
\begin{equation}
    H_{W,+} = X_W \sigma_x - Y_W \sigma_y + Z_W \sigma_z ,
\end{equation}
\begin{equation}
    H_{W,-} = X_W \sigma_x + Y_W \sigma_y + Z_W \sigma_z ,
\end{equation}
\begin{equation}
    \begin{split}
    X_W &= k_z - a k_y ~~,~~ Y_W = \chi k_y ,\\
    Z_W &= k_y^2 + k_z^2 - \Delta^2 .
    \end{split}
\end{equation}
The Berry connection of the occupied band of $H_{W,+}$ is calculated as
\begin{equation}
    \begin{split}
        A_{W,+,y} = \frac{-\chi (k_z + a k_y^2)}{2R_W (R_W - Z_W)} ,\\
        A_{W,+,z} = \frac{\chi k_y}{2R_W (R_W - Z_W)} ,
    \end{split}
\end{equation}
where $R_W$ is defined as
\begin{equation}
    R_W = \sqrt{ X_W^2 + Y_W^2 + Z_W^2} .
\end{equation}
The $x$ component of the Berry curvature is given as
\begin{equation}
    \begin{split}
    B_{W,+,x} =& \frac{\partial A_{W,+,z}}{\partial k_y} - \frac{\partial A_{W,+,y}}{\partial k_z} \\
    =& \frac{\chi}{2R_W^3} (Z_W + 2 \Delta^2 + 2 a k_y^2 k_z) .
    \end{split}
\end{equation}
The $chi$ dependence of $B_{W,+,x}$ is shown in Fig.\ref{fig:Wberry}. Similarly to the case of the L-point, the sharp peak feature in a small $\chi$ region is seen. The Chern number of the occupied band of the $+i$ block is calculated by integrating the Berry connection on a closed path $(k_y,k_z)=(k \cos \theta, k \sin \theta + a k^2 \cos^2 \theta)$,
\begin{equation}
    \begin{split}
    C_{W,+} =& \frac{1}{2\pi} \oint d \bm{k} \cdot \bm{A} \\
    =& \frac{1}{2\pi} \int_0^{2\pi} d \theta \frac{\chi k^2}{k^2 (\cos^2 \theta + \chi^2 \sin^2 \theta)} \\
    =& \frac{\chi}{| \chi |} .
    \end{split}
\end{equation}
Note that the closed path always involves the points where sharp peaks appear. The mirror Chern number $n_{W,{\cal M}_{(001)}}$ is given as
\begin{equation}
    \begin{split}
    n_{W,{\cal M}_{(001)}} =& \frac{C_{W,+}-C_{W,-}}{2} \\
    =& \frac{\chi}{| \chi |} = \left\{
    \begin{array}{ll}
        1 & (\chi>0) \\
        -1 & (\chi<0)
    \end{array} \right. .
    \end{split}
\end{equation}

\subsection{$\Sigma$-line}
The local model with SOC is written as
\begin{equation}
    \begin{split}
    H_{\Sigma,\mathrm{soc}} =& k_z \sigma_x s_0 + (k_x^2 - k_y^2 + k_z^2 + b) \sigma_z s_0 \\
    & + \chi \sigma_y (-k_y s_x + k_x s_y) .
    \end{split}
\end{equation}

On the $k_x=0$ plane (the (001) mirror plane), the model is block diagonalized with a unitary transformation $U_1=\sigma_0 i \frac{1}{\sqrt{2}} (s_x + s_z)$ as
\begin{equation}
    H_{\Sigma,+}^{(k_x=0)} = X_{\Sigma,x} \sigma_x + Y_{\Sigma,x} \sigma_y + Z_{\Sigma,x} \sigma_z ,
\end{equation}
\begin{equation}
    H_{\Sigma,-}^{(k_x=0)} = X_{\Sigma,x} \sigma_x - Y_{\Sigma,x} \sigma_y + Z_{\Sigma,x} \sigma_z ,
\end{equation}
\begin{equation}
    \begin{split}
    X_{\Sigma,x} &= k_z ~~,~~ Y_{\Sigma,x} = \chi k_y \\
    Z_{\Sigma,x} &= - k_y^2 + k_z^2 + b .
    \end{split}
\end{equation}
The mirror operator about the $k_x=0$ plane is also transformed as $U_1^\dagger (-i \sigma_0 s_x) U_1 = -i \sigma_0 s_z$. The Berry connection of the occupied band of $H_{\Sigma,+}^{(k_x=0)}$ is calculated as
\begin{equation}
    \begin{split}
        A_{\Sigma,+,y}^{(k_x=0)} &= \frac{\chi k_z}{2R_{\Sigma,(k_x=0)} (R_{\Sigma,(k_x=0)} - Z_{\Sigma,x})} , \\
        A_{\Sigma,+,z}^{(k_x=0)} &= \frac{-\chi k_y}{2R_{\Sigma,(k_x=0)} (R_{\Sigma,(k_x=0)} - Z_{\Sigma,x})} , \\
        R_{\Sigma,(k_x=0)} &= \sqrt{ X_{\Sigma,x}^2 + Y_{\Sigma,x}^2 + Z_{\Sigma,x}^2}
    \end{split}
\end{equation}
and the $x$ component of the Berry curvature is given as
\begin{equation}
    \begin{split}
    B_{\Sigma,+,x}^{(k_x=0)} =& \frac{\partial A_{\Sigma,+,x}^{(k_x=0)}}{\partial k_y} - \frac{\partial A_{\Sigma,+,x}^{(k_x=0)}}{\partial k_z} \\
    =& \frac{-\chi}{2R_{\Sigma,(k_x=0)}^3} (Z_{\Sigma,x} -2b) .
    \end{split}
\end{equation}

On the $k_y=0$ plane (the (110) mirror plane), the model is block diagonalized with a unitary transformation $U_2=\sigma_0 i \frac{1}{\sqrt{2}} (s_y + s_z)$ as
\begin{equation}
    H_{\Sigma,+}^{(k_y=0)} = X_{\Sigma,y} \sigma_x + Y_{\Sigma,y} \sigma_y + Z_{\Sigma,y} \sigma_z ,
\end{equation}
\begin{equation}
    H_{\Sigma,-}^{(k_y=0)} = X_{\Sigma,y} \sigma_x - Y_{\Sigma,y} \sigma_y + Z_{\Sigma,y} \sigma_z ,
\end{equation}
\begin{equation}
    \begin{split}
    X_{\Sigma,y} &= k_z ~~,~~ Y_{\Sigma,y} = \chi k_y \\
    Z_{\Sigma,y} &= k_x^2 + k_z^2 + b .
    \end{split}
\end{equation}
The mirror operator about the $k_y=0$ plane is also transformed as $U_2^\dagger i \sigma_0 s_y U_2 = i \sigma_0 s_z$. The Berry connection of the occupied band of $H_{\Sigma,+}^{(k_y=0)}$ is calculated as
\begin{equation}
    \begin{split}
        A_{\Sigma,+,x}^{(k_y=0)} &= \frac{\chi k_z}{2R_{\Sigma,(k_y=0)} (R_{\Sigma,(k_y=0)} - Z_{\Sigma,y})} , \\
        A_{\Sigma,+,z}^{(k_y=0)} &= \frac{-\chi k_y}{2R_{\Sigma,(k_y=0)} (R_{\Sigma,(k_y=0)} - Z_{\Sigma,y})} , \\
        R_{\Sigma,(k_y=0)} &= \sqrt{ X_{\Sigma,y}^2 + Y_{\Sigma,y}^2 + Z_{\Sigma,y}^2}
    \end{split}
\end{equation}
and the $y$ component of the Berry curvature is given as
\begin{equation}
    \begin{split}
    B_{\Sigma,+,y}^{(k_y=0)} &= \frac{\partial A_{\Sigma,+,x}^{(k_y=0)}}{\partial k_z} - \frac{\partial A_{\Sigma,+,z}^{(k_y=0)}}{\partial k_x} \\
    &= \frac{\chi}{2R_{\Sigma,(k_y=0)}^3} (Z_{\Sigma,y} -2b) .
    \end{split}
\end{equation}

The mirror Chern numbers are numerically calculated and the result is
\begin{equation}
    \begin{split}
       n_{\Sigma,{\cal M}_{(001)}} = \left\{
       \begin{array}{ll}
            1 & (b>0) \\
            0 & (b<0)
       \end{array}
       \right. , \\
       n_{\Sigma,{\cal M}_{(110)}} = \left\{
       \begin{array}{ll}
            0 & (b>0) \\
            1 & (b<0)
       \end{array}
       \right. ,
    \end{split}
\end{equation}
as shown in the main article.

\section{Material Examples}
 Here we give the results of the first-principles calculations in FCC Ca and Ba. In addition to the figure in main article, we show the irreps of the occupied band.
 
In Ca, we need to check only a band around $-4$ eV $\sim$ $0$ eV as an occupied band. The irreps of the band are shown in Table \ref{tab:Cairreps}.

\begin{table}[]
    \centering
    \begin{tabular}{|c||c|c|c|c|}
    \hline
        (Ca) & $\Gamma$ & X & L & W  \\ \hline
        w/o SOC & $\Gamma_1^+$ & $\Gamma_1^+$ & $\Gamma_1^+$ & $\Gamma_4$ \\ \hline
        w/ SOC & $\Gamma_6^+$ & $\Gamma_6^+$ & $\Gamma_4^+$ & $\Gamma_7$ \\ \hline
    \end{tabular}
    \caption{Irreps of the occupied band of FCC Ca.}
    \label{tab:Cairreps}
\end{table}

In Ba, we need to check only a band around $-3$ eV $\sim$ $0$ eV as an occupied band. The irreps of the band are shown in Table \ref{tab:Bairreps}.

\begin{table}[]
    \centering
    \begin{tabular}{|c||c|c|c|c|}
    \hline
        (Ba) & $\Gamma$ & X & L & W  \\ \hline
        w/o SOC & $\Gamma_1^+$ & $\Gamma_1^+$ & $\Gamma_1^+$ & $\Gamma_4$ \\ \hline
        w/ SOC & $\Gamma_6^+$ & $\Gamma_6^+$ & $\Gamma_4^+$ & $\Gamma_7$ \\ \hline
    \end{tabular}
    \caption{Irreps of the occupied band of FCC Ba.}
    \label{tab:Bairreps}
\end{table}

\end{document}